\documentclass[useAMS,usenatbib]{mn2e}
\usepackage{epsfig}
\usepackage{multirow}
\usepackage{color}

\newcommand{\teff}{$T_{e\!f\!f}$}
\newcommand{\kms}{km s$^{-1}$}
\newcommand{\logg}{$\log g$}

\title[ARGOS bulge survey]{ARGOS II: The Galactic Bulge Survey}
\author[K. Freeman et al.]
  {K.~Freeman$^1$,\thanks{E--mail:kcf@mso.anu.edu.au.}
  M.~Ness$^1$, E.~Wylie--de--Boer$^1$, E.~Athanassoula$^2$, 
  J.~Bland--Hawthorn$^3$, 
  \newauthor 
 M.~Asplund$^1$,
 G.~Lewis$^3$,
  D.~Yong$^1$,
 R.Lane$^4$,
  L.~Kiss$^{3,5,7}$,
  and R.~Ibata,$^6$  \\
$^1$Research School of Astronomy \& Astrophysics, Australian National University, Cotter Rd., Weston, ACT 2611, Australia\\
$^2$Aix Marseille Universit\'e, CNRS, LAM (Laboratoire d'Astrophysique de
Marseille) UMR 7326, 13388, Marseille, France  \\
$^3$Sydney Institute for Astronomy, University of Sydney, School of Physics A28, NSW 2006, Australia\\
$^4$Departamento de Astronom\'{i}a Universidad de Concepci\'{o}n, Casilla 160 C, Concepci\'{o}n, Chile\\ 
$^5$MTA CFSK, Konkoly Observatory, Konkoly Th. M. 15-17, H-1121, Hungary. \\ 
$^6$Observatoire Astronomique, Universit\'{e} de Strasbourg, CNRS, 11 rue de l'Universit\'{e}, F--67000,      Strasbourg, France \\
$^7$ELTE Gothard-Lend\"ulet Research Group, H-9700, Szombathely, Hungary.\\
 } 

\begin{document}

\date{Accepted 2012 TBD. Received 2012. As soon as possible; in original form 2012 August}

\pagerange{\pageref{firstpage}--\pageref{lastpage}} \pubyear{2012}

\maketitle

\label{firstpage}

\begin{abstract}
We describe the motivation, field locations and stellar selection for the ARGOS spectroscopic survey of 28,000 stars in the bulge and inner disk of the Milky Way galaxy across latitudes of $b=-5^\circ$ to $-10^\circ$. The primary goal of this survey is to constrain the formation processes of the bulge and establish whether it is predominantly a merger or instability remnant. From the spectra (R = 11,000), we have measured radial velocities and determined stellar parameters, including metallicities and [$\alpha$/Fe] ratios. Distances were estimated from the derived stellar parameters and about 14,000 stars are red giants within 3.5 kpc of the Galactic centre. In this paper, we present the observations and analysis methods. Subsequent papers (III and IV) will discuss the stellar metallicity distribution and kinematics of the Galactic bulge and inner disk, and the implications for the formation of the bulge.  
\end{abstract}


\section{Introduction}
 
The Abundances and Radial velocity Galactic Origins Survey (ARGOS) is a large spectroscopic survey of about 28,000 stars in the Galactic bulge and surrounding disk, using the AAOmega fibre spectrometer on the Anglo-Australian Telescope.  With this survey, we aim to investigate the metallicities, kinematics and density distributions of the stars in the inner region and use this information to understand the bulge. Near-infrared images of the Galactic bulge (e.g. the COBE/DIRBE images) show its peanut/boxy structure.  Peanut/boxy bulges are widely believed to be generated via bar-forming and bar-buckling instabilities of the disk, and  the main motivation of this survey is to understand how the Galactic bulge formed. Did it form primarily through disk instability or did mergers play a significant role in its formation? 

Several authors \citep[e.g.][]{Athanassoula2005, Inma2006, Shen2010} have made N-body simulations of peanut/boxy bulge formation through bar-forming and bar-buckling instability of the disk.  One of the goals of the ARGOS survey is to see whether the existing N-body models provide a good representation of the bulge properties, in terms of the observed kinematics and peanut/boxy structure.  The ARGOS data provide 3D positions and radial velocities for the stars, and also their [Fe/H] and [$\alpha$/Fe] abundances. The combination of positional, kinematical and chemical data will be useful for testing future models of bulge formation that include chemical evolution.   

A generic prediction of the N-body models is that instability-generated bulges are expected to show cylindrical rotation, unlike the rotation field predicted for the more spheroidal classical bulges. The prediction of cylindrical rotation for the boxy bar/bulge is easily checked from radial velocity surveys of the bulge, and has already been confirmed by the BRAVA group \citep{Howard2009}.  Testing for the presence of a classical component of the bulge is more difficult. \citet{Saha2012} showed that a small classical bulge in the presence of a boxy bar/bulge would itself be spun up into cylindrical rotation and would therefore lose its distinct kinematical signature.  

Recently, a split red clump in the Galactic bulge has been discovered \citep[see][]{Saito2011,Nataf2010,McWilliam2010}. The split is seen as a bimodal magnitude distribution of the red clump stars at higher latitudes near the minor axis of the bulge, and is associated with the boxy/peanut structure of the bulge. In projection, the boxy/peanut bulge shows an underlying X-structure.  Our data are well suited to investigating the stellar populations present in the X-structure. The first paper of this series is on the X-structure observed in the bulge \citep{Ness2012}, and showed that this characteristic structure of the bulge is defined by stars with metallicities [Fe/H] $> -0.5$.  The more metal-poor stars do not appear to show the split red clump and are therefore probably not involved in the boxy/peanut structure.  

If the Galactic bulge formed from instability of the inner disk many Gyr ago, then the bulge would be made up mainly of early thin and thick disk stars that have been trapped within the bulge structure.   These stars would present as a fossil of the early inner Galaxy, and we could expect to see evidence of these trapped early components in the metallicity distribution and kinematics of the bulge.  The relative contributions of the early components are likely to vary across the bulge, and this variation may give insight into the mapping of the early disk into the bulge via the instability.  We therefore chose to survey stars in the bulge and also in the surrounding thin and thick disk, to see whether the metallicity distribution across the inner regions of the Galaxy is consistent with the instability picture. In this context, the Galactic bulge appears to show a vertical abundance gradient \citep[e.g.][]{Minniti1995, Zoccali2008, Babusiaux2010} which could be interpreted as evidence for a dissipative formation.  It is of interest to investigate this abundance gradient in more detail. Futhermore, the [$\alpha$/Fe] properties of the bulge have been much discussed \citep[e.g][]{Fulbright2007, Zoccali2007, Melendez2008, AlvesBrito2010, Johnson2011}, with indications from some authors that even the most metal-rich stars in the bulge are $\alpha-$enhanced. \citet{Bensby2010} also find evidence for enhanced [$\alpha$/Fe] in a limited sample of inner disk giants (typically $>$ 1 kpc above the plane). The [$\alpha$/Fe] ratios for stars in the inner Galaxy are an important diagnostic of the rate of star formation in this region. We therefore wanted to measure the [$\alpha$/Fe] ratios for the large sample of ARGOS bulge stars. 

There is a possibility that the first stars are concentrated to the inner regions of the Galaxy, as indicated in cosmological simulations \citep[e.g][]{Diemand2005, Brook2007}. These stars are of great interest for understanding the early chemical evolution of the Galactic environment. The selection criteria for the survey have been defined to ensure we did not exclude such metal-poor giants. 

Along lines of sight passing through the inner region of the Galaxy are stars from all the major Galactic components: thin and thick disk, bulge and stellar halo. Some of these are likely to be involved in the peanut/boxy bulge structure and some are not.  Contamination by foreground and background disk stars is an issue for surveys of stars in the Galactic bulge. Galactic models like Galaxia \citep{Sharma2011} show that samples of red giants in the direction of the bulge can include significant fractions of stars of the thin and thick disk lying in front of and behind the bulge.  Stars of the inner disk are of interest for understanding how the bulge relates to its environment, particularly if it did form via disk instability. For specific studies of metallicities and kinematics in the bulge, however, we want to choose stars such as red clump giants for which accurate ($\sim 10$\%) distances can be estimated.  With these distances, we can then determine which stars lie within the bulge and identify the stars which are in the foreground and background and derive a relatively uncontaminated picture of the distribution of kinematics and chemical properties throughout the bulge.  For example, with the selection criteria described below, designed to select mainly red clump giants along sight lines through the bulge, we found that about 2,500 of our 28,000 stars are foreground dwarfs and about 14,000 stars lie within the bulge region (Galactocentric cylindrical radius $R_G < 3.5$ kpc). The remaining 11,500 stars are in the foreground or background and outside the inner region.

The overall outcome of the survey is that the bulge appears to be consistent with formation via bar forming and bar buckling instability of the early disk. The dominant component of the peanut/boxy bulge is the early old thin disk, with a clear contribution from its more metal-rich and kinematically colder component, similar to the younger, more metal-rich and colder component seen in the thin disk near the Sun at the present time \citep[e.g.][]{Haywood2008}.

The first paper of this series \citep{Ness2012} showed that the peanut/boxy structure of the bulge is defined by stars with metallicities [Fe/H] $> -0.5$, and its structure and kinematics are well represented by N-body models of bulge formation via disk instability. This paper, ARGOS II, deals with the observational methods and strategy for the ARGOS survey. Paper III will present the components of the metallicity distribution for the bulge, and paper IV presents the kinematics of its sub-populations.  Paper V will discuss the mapping of the early disk into the bulge by the instability.

\section{The Data}

In this section, we describe the selection criteria for the stars, and provide details of the AAOmega observations.

\subsection{Sample Selection}

\begin{figure}
\begin{center}
 \includegraphics[width=0.45\textwidth]{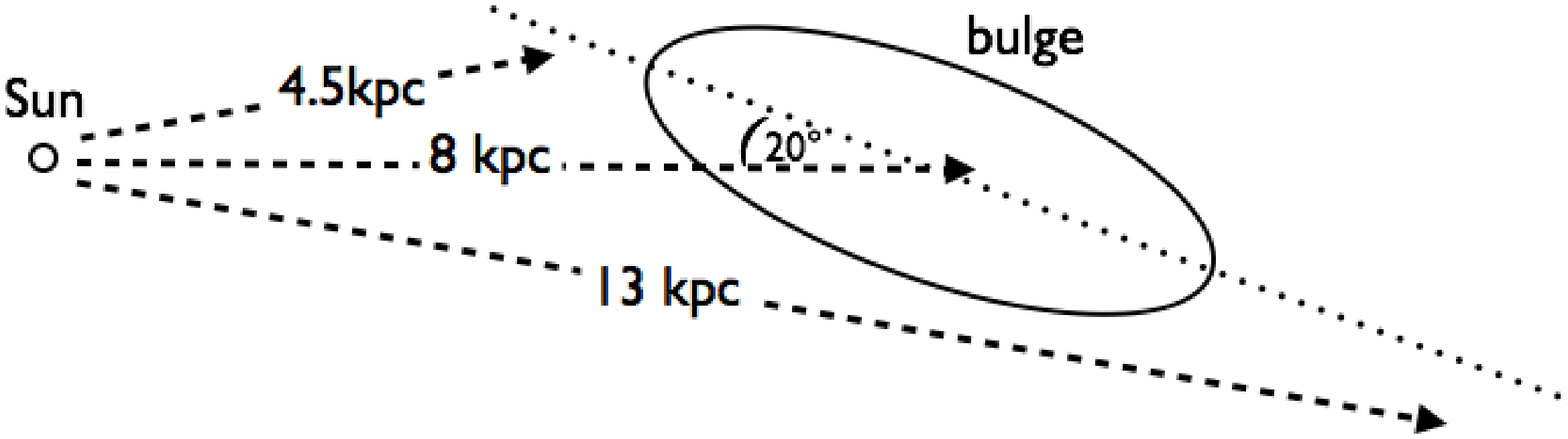} 
\caption{This sketch in the $xy$-plane illustrates the range of distance from the Sun that we wish to cover in each field with our selection of clump giants.}
\label{fig:cartoon3}
\end{center}
\end{figure}

Our goal is to select red giants in a grid of fields over the bulge and inner disk, with an apparent magnitude and colour range that includes clump giant stars all along the line of sight through the bulge, from the near side to the far side, and surrounding inner disk. The distance range is about $4.5$ to $13$ kpc from the Sun (see Figure \ref{fig:cartoon3}).  We also wish to include metal-poor giants in the inner Galaxy, as described above, because of their interest as possible first stars.  With the above goals and criteria in mind, we minimise the numbers of the brightest and coolest giants for which abundances are difficult to measure reliably from optical spectra. Foreground bluer main sequence stars should be excluded without excluding metal-poor giants along the line of sight.

\begin{figure}
\centering
 \includegraphics[width=0.40\textwidth]{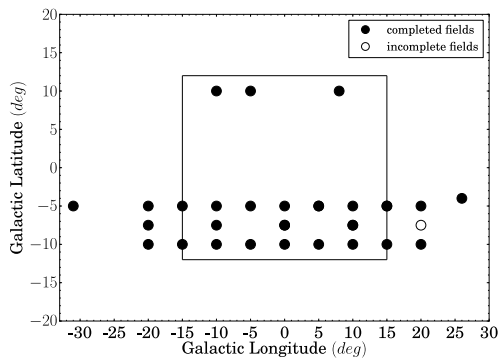}
\caption{Galactic  latitude and longitude for the 28  $2^\circ$ fields in our survey.  The rectangle shows the approximate extent of the boxy bulge. The filled circles are fields for which about $1000$ stars were observed. The open circle is our one incompletely observed field ($\sim$ 600 stars).}
\label{fig:donefields}
\end{figure}

In order to complete the survey in a realistic time, we used the sparse sampling shown in Figure \ref{fig:donefields}. This grid of field centers  covers the southern bulge between $b = -5^\circ$ to $-10^\circ$, extending out in longitude into the surrounding thin disk and thick disk. For the stars in the bulge region, we choose field centres at latitudes $|b| \ge 5^\circ$, because the mean reddening increases rapidly closer to the Galactic plane. The field centres also include two low-latitude fields in the disk at larger longitudes in windows of low reddening, and three relatively low extinction fields in the northern bulge for comparison with the southern fields as a check on north-south symmetry. About 1000 stars were observed in each field. Stars were selected from the Two Micron All Sky Survey (2MASS) catalogue \citep{2MASS} over the magnitude range K = 11.5 to 14.0, with J,K errors $< 0.06$, and flags set to reduce blends, contamination, stars of lower photometric quality and non-stellar objects (Qflg=AAA, Bflg=111, Cflg=000, Xflg=0, Aflg=0).  

\begin{figure}
          \includegraphics[width=0.45\textwidth]{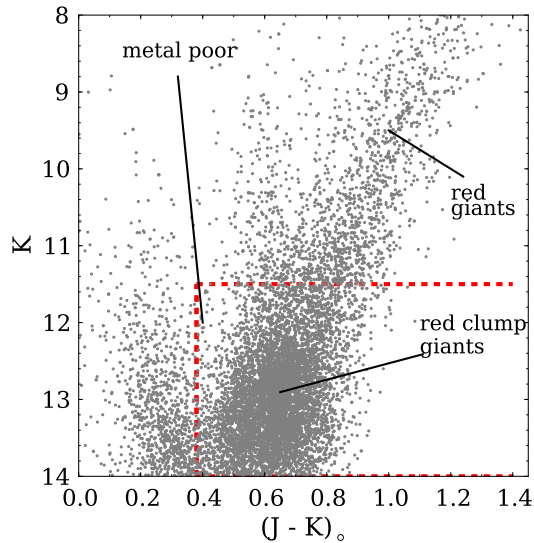} 
 \caption{Colour Magnitude Diagram from 2MASS for the bulge field at $l=-10^\circ, b=-5^\circ$. The red giant clump at the distance of the Galactic centre is near $(J-K)_o = 0.65, K = 13.2$. The broken lines show the region from which our stars are selected. This region includes mainly red clump giants along the line of sight through the bulge, plus some brighter and fainter giants lying mostly on the far and near side of the bulge respectively.  The plume of foreground disk stars to the left of the CMD is excluded by the colour cut in $(J-K)_o$ but metal-poor giants (i.e. [Fe/H] $<$ --2.0) are included as shown.}
   \label{fig:singlefield}
\end{figure}

The red clump giants have a well-defined absolute magnitude distribution, with a mean $ M_{K_0}$ = --1.61 $\pm$ 0.03 \citep{Alves2000} and a standard deviation of $0.22$ mag. This allows a relatively accurate estimate of their distances (standard error about $0.9$ kpc at a distance of $8$ kpc). The clump giants lie in the plume near $(J-K)_o = 0.65$ in Figure \ref{fig:singlefield}. With the range of apparent magnitudes selected for our survey, the clump stars are distributed over the whole line of sight through the bulge (Figure \ref{fig:cartoon3}). Some intrinsically brighter/fainter giants on the far/near side of the bulge are included in the sample, although the magnitude cut at $K = 11.5$ excludes most of the brighter and cooler giants for which it is more difficult to measure abundances.  A blue colour cut was made to exclude the plume of bluer foreground stars seen in Figure \ref{fig:singlefield}. The colour of this cut is at $(J-K)_o$ = 0.38 so depends on the mean reddening in the field. The stars in the sample are then mainly clump giants, plus a contribution from brighter and fainter giants and some metal-poor giants which come mainly from the inner regions of the metal-poor halo of the Galaxy.  We can expect some contamination from nearby G and K dwarfs, but they can be easily removed using their inferred stellar parameters.

\begin{figure}
 \centering
  \includegraphics[width=0.50\textwidth]{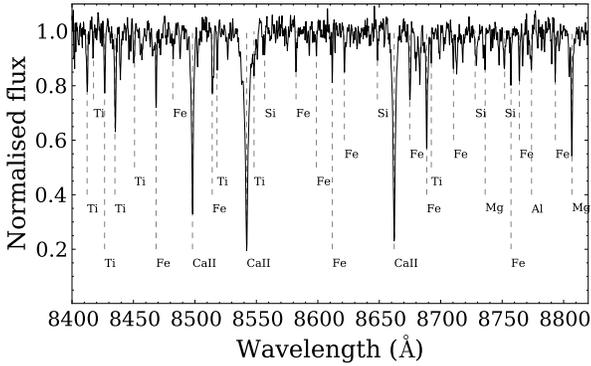}
        \caption{ A typical AAOmega stellar spectrum of a red giant star in the Ca-triplet region after continuum normalisation. Lines of some elements are indicated. All species are singularly ionised except Ca. The star shown here has stellar parameters of \teff\ = 4600 K, \logg\ = 2.8, [Fe/H] = 0 and a S/N = 70 per resolution element.}
   \label{fig:spectrastar}
\end{figure}

The stellar parameters are derived from our primary wavelength region around the Ca-triplet. The highest spectral resolution available with the AAOmega system \citep{Sharp2006} at the Ca-triplet region (R = 11,000) is well suited to our goals: it provides accurate radial velocities and abundances for a few individual elements including [Fe/H] and $\alpha$-elements (see Figure \ref{fig:spectrastar}).  The stars are sufficiently bright to observe in bright time. 

The mean \citet{Schlegel} reddening for each ARGOS field was adopted to make the blue cut.  The reddening ranges from $E(J-K) = 0.06$ to $0.17$ in the $b = -10^\circ$ fields, $0.12$ to $0.24$ at $b = -7.5^\circ$,  $0.24$ to $0.39$ at $b = -5^\circ$ and $0.12$ to $0.21$ in the northern fields and disk well away from the bulge. Most of our stars lie at $z$-heights above $300$ pc, so the \citet{Schlegel} reddening is likely to be a realistic estimate. We have used the \citet{Schlegel} reddening throughout this study, which was completed before the \citet{Gonzalez2011} reddening map for the bulge became available. The \citet{Gonzalez2011} reddening in our field at $(l,b) = (0^\circ,-5^\circ)$ is about $0.1$ lower in $E(J-K)$ than the \citet{Schlegel} reddening. The observed colour of the low density region in the CMD at $(J-K)_o = 0.38$, where the foreground star plume and the clump star sequence meet, provides a check on the mean reddening in the field. In all fields, its colour was consistent with the \citet{Schlegel} reddening. The blue cut is sufficiently blue such that metal-poor giants in the field are not excluded.

Towards the faint end of the selection interval in $K$-magnitude, the 2MASS catalog in the bulge fields becomes rapidly incomplete. We wanted to have clump stars all along the line of sight through the bulge, so chose roughly equal numbers of stars per unit magnitude in each field. It was also desirable to have a relatively small magnitude range for each fibre setup to reduce the effects of fibre-to-fibre scattering. Because the Ca-triplet region is close to the photometric $I$-band, we estimated the I-magnitude for each star from its $JK$ photometry, using the transformation  $I_o = J_o + 1.095\,(J-K) + 0.421\,E(B-V)$, derived from \citet{BessellBrett1988} and adopting the ratios  $E(J-K)/E(B-V) = 0.535$, $A_K = 0.347\,E(B-V)$ and $A_{I}/A_{V}$ = 0.601 and reddening values \citep{Schlegel} taken from the NASA/IPAC Extragalactic Database (NED). Each field typically had many thousands of stars satisfying the selection criteria in Figure \ref{fig:singlefield}.

\subsection{Observations}
The spectra were acquired with the fibre-fed AAOmega system on the Anglo-Australian Telescope in May-August of 2008 to 2011.  AAOmega is a dual-beam spectrometer with about 400 fibres that can be positioned within a $2-$degree diameter field.  With AAOmega (allowing for broken fibres), we can observe about 350 stars at once, plus 25 sky positions and guide fibres. The red beam provides our primary spectral region, using the 1700D grating with a resolution of about 11,000 in the Ca-triplet region $8400 - 8850$\AA. This region is relatively free from telluric lines and contains spectral lines of  Fe I, Ti I, Ca II, Al I, Mg I, Si I and O I for abundance measurements: see Figure \ref{fig:spectrastar}.  Of the optical regions of the spectrum that we could have chosen for this project, the Ca-triplet region is least affected by moonlight and it was possible to conduct the program during bright lunar phase. On the other hand, the region is not ideally suited to estimating the spectroscopic stellar surface gravities which are needed to determine which stars are clump giants and to estimate the distances for red giants that are not on the clump. The typical signal to noise ratio (S/N) for our Ca-triplet spectra is about $75$ per resolution element.  The blue beam was used to acquire low resolution spectra (R $\sim$ 3000) with the 1500V grating in the region $5000 - 5600$\AA. The main purpose of these low resolution spectra is to reject foreground dwarf stars from our sample of more distant giants, using the Mgb/MgH feature.

The spectra from the fibres in each of the red and blue beams are imaged on to a 2K $\times$ 2K E2V CCD. For each field, $25$ fibres were used to measure the sky spectrum.   Wavelength calibration spectra were acquired before and after the exposures for each fibre setup, and flat field spectra were taken for each setup in order to trace the spectra of the stars and to remove small fringing effects and pixel-to-pixel sensitivity variations. 

Figure \ref{fig:donefields} shows the locations of the 28 fields selected for this program. Spectra were acquired for about 28,000 stars, using 42 nights in the winters of $2008-2011$.  Our goal was to acquire spectra of about $1000$ stars in each field. This was achieved, except for a single field at $(l,b) = (20^\circ,-7.5^\circ)$ for which we have about $600$ stars.

 The $I-$magnitudes of the selected stars were mostly between about $13$ and $16$ and were used to allocate each star to one of three one-magnitude intervals for observation. These magnitude intervals were approximately from $I = 13-14, 14-15$ and $15-16$, depending on the mean reddening in the individual fields.  In each field, we selected about $1000$ stars, uniformly distributed over the three magnitude intervals and uniformly distributed over the $2^\circ$ field so that every available fibre could be allocated to a star.  The typical total exposure times for the three magnitude intervals were (60, 120, 150) minutes respectively, acquired in several sub-exposures, so the S/N is usually higher for the brighter stars. The clump stars in these $I-$band intervals are at mean distances of about ($5.5, 8.5, 10.5$) kpc respectively, covering the entire line of sight through the bar/bulge at longitudes between $+20$ and $-20$ degrees. For each field, $25$ sky positions, which are free of visible stars on the red Digital Sky Survey survey images, were selected visually. 

The S/N achieved in each field for our primary spectrum over the Ca-triplet region is given in Table \ref{table:one}, where the different magnitude ranges are indicated using 1, 2, 3, from the brightest to the faintest. The code for the field names is the longitude and then the latitude, where the m and p are for minus and plus values of $l$ and $b$: e.g. m0m10\_3 denotes the fibre setup  at $(l, b) = (0^\circ,-10^\circ)$ and at the faintest magnitude range. 

\begin{table}
\centering
\caption{Mean S/N per resolution element in the Ca-triplet region 
for each field.}
 \begin{tabular}[]{| l | l | l | l | l | l |}
\hline
Field Name & S/N& Field Name & S/N &  Field Name&S/N   \\
     \hline
m0m10$\_$1 & 93  &  m0m10$\_2$ & 91&  m0m10$\_3$ & 87 \\
m0m5$\_1$ & 80  & m0m5$\_2$& 86 & m0m5$\_3$& 76 \\
m5m5$\_1$ &72 &m5m5$\_2 $&74 &m5m5$\_3 $&64\\
m5p10$\_1$ &91 & m5p10$\_2$& 77 & m5p10$\_3$& 80\\
m5m10$\_1$ &88 & m5m10$\_2$& 78  &m5m10$\_3$& 82\\
m10m10$\_1$ &74  & m10m10$\_2$ &55&   m10m10$\_3$& 82 \\
m10m5$\_1$ & 70  & m10m5$\_2$ & 56  &  m10m5$\_3 $& 69 \\
m10m75$\_1$& 84  & m10m75$\_2$ & 82 &  m10m75$\_3$ &68 \\ 
m10p10$\_1$ & 89 & m10p10$\_2$ & 68 & m10p10$\_3$ & 85  \\
m15m10$\_1$ &114 &m15m10$\_2$&73 & m15m10$\_3$& 60 \\
m15m5$\_1$ & 99 &m15m5$\_2 $&77 & m15m5$\_3$& 75 \\
m20m10$\_1$& 95 &  m20m10$\_2$& 72 & m20m10$\_3$& 63\\
m20m5$\_1 $&72   &m20m5$\_2$ &84 & m20m5$\_3$ &74 \\
m20m75$\_1$& 95 &  m20m75$\_2$ &80   & m20m75$\_3 $&76\\ 
m31m5$\_1 $&80   & m31m5$\_2 $&90 & m31m5$\_3 $&64\\
m5m5$\_1 $&72 & m5m5$\_2$& 74 & m5m5$\_3$ &44\\
m5p10$\_1$ &92 & m5p10$\_2 $&77 & m5p10$\_3 $&80\\ 
p10m10$\_1$& 122 & p10m10$\_2$ &85 & p10m10$\_3$& 74\\ 
p10m75$\_1$ &79& p10m75$\_2$& 42 & p10m75$\_3$ &  41 \\
p10m5$\_1 $&66 & p10m5$\_2$& 89 & p10m5$\_3$ &74\\ 
p15m10$\_1$ & 64 & p15m10$\_2$ &81 & p15m10$\_3$& 75\\ 
p5m10$\_1 $& 108 & p5m10$\_2$& 67 &p5m10$\_3$& 37\\
p5m5$\_1$ & 90 & p5m5$\_2$& 61&pm5m$\_3$ & 52\\
p26m4$\_1$ & 73 & p26m4$\_2$ &73& p26m4$\_3$ &77\\ 
p20m10$\_1$ & 98&  p20m10$\_2$ &60 & p20m10$\_3$& 104\\
p20m5$\_1 $& 52  & p20m5$\_2$& 89 & p20m5$\_3$& 59\\ 
p20m75$\_1$ & 77 & p20m75$\_2$ &44 & &\\
p26m4$\_1 $& 73 &  p26m4$\_2$ & 73 &  p26m4$\_3$ & 77 \\
p5m10$\_1 $& 107 & p5m10$\_ 2$  & 67 & p5m10$\_3$ & 110 \\
p8p10$\_1 $& 110 & p8p10$\_2$ & 75 & p8p10$\_3$  & 80\\
p15m10$\_1$ & 64 & p15m10$\_2$ & 81 & p15m10$\_3$ & 68\\
m0m75$\_1 $&  88 & m0m75$\_2$ & 55  &m0m75$\_3$ &  64 \\
\hline
\end{tabular}
\label{table:one}
\end{table}

\section{Data Reduction}

The first stage of the reduction used the standard 2dfdr pipeline\footnote{www2.aao.gov.au/twiki/bin/view/Main/CookBook2dfdr }. This program subtracts bias and scattered light, extracts the individual stellar spectra, corrects for the flat field, calibrates wavelength and fibre throughput and subtracts sky.  We selected the optimal sky subtraction option which gives much better sky subtraction, and used a 4th order polynomial for fitting the arc lamp lines. 

The automated sky subtraction did not work well for low S/N data or on nights that were partially cloudy.  Therefore, the sky subtraction was done externally with a python program that subtracted the sky on a star by star basis using the sky emission lines for normalizing the throughputs of the star and sky fibres. The continuum sky was measured from the dedicated sky fibres; this improved the sky subtraction and reduced the internal error of the radial velocity measurements via cross correlation with synthetic template spectra. For estimating the stellar parameters,  the wavelength intervals of the strong sky lines were excluded from the analysis to ensure that the estimates were not affected by errors in sky subtraction. This turned out to be an unnecessary precaution, because the parameter estimates were unchanged whether or not the regions of the strong sky lines were included.  The derived stellar parameters were similar for spectra reduced entirely with 2dfdr and spectra for which the external sky subtraction was used, except for the effect on the error of the radial velocities mentioned above.

Scattered light removal is critical to ensure the correct continuum level.  The  scattered light removal with 2dfdr was checked using independent IRAF ({\it dofiber}) reductions.  The IRAF results were consistent with those from 2dfdr. 

\begin{figure}
\centering
   \includegraphics[width=0.50\textwidth]{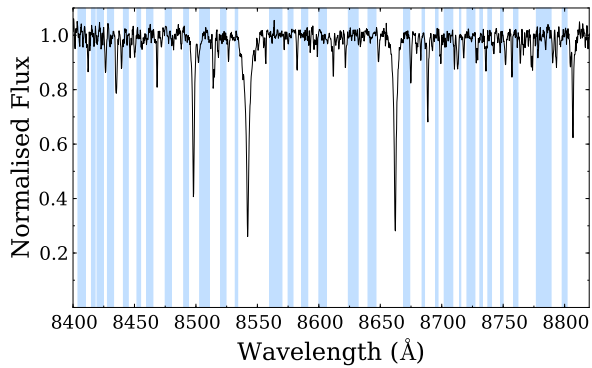}
 \caption{Spectrum of a representative star from Figure 2 showing the continuum regions in the blue shaded areas, identified using the \citet{Hinkle2005} atlas, used for normalisation of the AAOmega spectra}
   \label{fig:continuum}
\end{figure}

After data reduction, the spectra were normalized by fitting either a low order spline or Legendre polynomial to the median values of continuum regions identified from the Hinkle Arcturus Atlas (see Figure \ref{fig:continuum}).  A small fraction of the spectra were not well fit using a single spline function; for these spectra we tried both functions (a 4th order Legendre polynomial and a 3rd order spline) and the better one was adopted.\\  

\section{Analysis}

All spectra were inspected by eye for an initial culling of the data to remove spectra that are significantly damaged by bad pixels or cosmic rays in critical regions and to remove double lined binaries.  This reduced the original sample to about 26,000 stars which were passed through the $\chi^2$ program and analysed. A further $500$ stars with S/N $< 35$ (per resolution element) were removed from the analysis, reducing the final sample to $25,500$ stars with a mean S/N of about $75$ $\pm$ $20$. The foreground cool dwarfs were eliminated from the sample using the $\chi^2$ program described below to estimate their gravities; this step follows a preliminary cut using the gravity-sensitive MgH and Mgb features in the blue region of the spectra to identify dwarfs. 

\subsection[]{Overview: Measuring Stellar Parameters}

Stellar parameters (\logg, [Fe/H] and [$\alpha$/Fe]) were determined using a minimum $\chi^2$ program written for this dataset. The program compares each observed spectrum to a large grid of synthetic spectra, generated using the LTE stellar synthesis program MOOG 2010 \citep[][and subsequent updates]{Sneden1973} from the \citet{Castelli2001} 1D model atmospheres with no convective overshooting \citep{Castelli2004}.  The grid covers the parameter space \teff\ $=  3750 - 6250$ K, [Fe/H] $= -5.0$ to $1.0$, $\log g$ $ = 0.0$ to $5.0$ and [$\alpha$/Fe] $ = 0.0$ to $+1.0$. The grid steps are the steps of the original opacity distribution functions ($250$ K in \teff, $0.5$ in \logg, $0.5$ in [Fe/H] and $0.10$ dex in [$\alpha$/Fe]), for a microturbulent velocity of $2.0$ \kms. An assumption of a constant microturbulent velocity may be a source of error in our analysis. However, given we are observing mostly red clump stars which have a similar \logg, any error introduced by this adopted value of $2.0$ \kms\ should be systematic. 

The temperature is not estimated by the minimum $\chi^2$ procedure because of the usual degeneracies involved in attempting to measure all of the stellar parameters from spectra in the Ca-triplet region. The temperature is determined from the $(J-K)_\circ$ colours of our giant stars using the calibration by \citet*{Bessell1998}. The temperature error, estimated to be $\pm 100$ K, comes from uncertainties in  (i) the interstellar reddening, and (ii) the 2MASS apparent magnitude.  All of our selected stars have $JHK$ errors $< 0.06$ mag.  Comparison with a set of calibration stars from the literature, presented in the Appendix, shows that our mean values of abundance and \logg\ are close to those of the calibration stars.  If we had adopted the  \cite{Alonso1999} empirical temperature - colour calibration, the temperatures would be lower by about 100 K in the mean (up to 200 K for the most metal-poor stars and zero for the more metal-rich stars with [Fe/H] $> 0.0$ for which the temperatures from the two calibrations agree). The mean change in temperature is within the errors of the calibration and would typically shift the $\chi^2$ [Fe/H] abundances down by about $0.1$ dex in the mean and up to about $0.2$ dex for the most metal-poor stars. 

The calibration stars include 16 bulge stars with [Fe/H] between $0.1$ and $-1.2$ observed by \citet{Zoccali2008} at high resolution, and open and globular cluster stars observed by Da Costa et al. (2011, private communication) and \citet{Lane2011} with the same instrument and grating. The clusters include the open cluster Melotte 66 with [Fe/H] $= -0.3$, and 6 globular clusters: 47 Tuc ([Fe/H] $= -0.75$), NGC280 ([Fe/H] $= -1.3$), NGC1904 ([Fe/H] $= -1.6$), NGC6658 ([Fe/H] =$ -1.7$), NGC2290 ([Fe/H] $= -1.9$) and NGC5034 ([Fe/H] $= -2.1$). The total range in [Fe/H] for the calibration stars is $0.1$ to $-2.1$ \footnote{ References for calibration stars:
47 Tuc: \citet{McWilliam2008},
NGC 288: \citet{Shetrone2000}, NGC 1904: \citet{Gratton1989, Kirby2008}, 
Melotte 66: \citet{Gratton1994,Sestito2008}
NGC 2298: \citet{McWilliam1992}, Arcturus: \citep{Ramirez2011}} .

The solar and Arcturus spectra from \citet{Hinkle2005} were also included in the calibration, after convolution to our resolution of 11,000. The errors in our parameters, as estimated from comparison of the $\chi^2$ program results with the calibration star parameters, are $\pm 100$ K in \teff,  $\pm 0.30$ in \logg,  $\pm 0.09$ in [Fe/H] and $\pm 0.10$ in [$\alpha$/Fe]. The comparison of the $\chi^2$ and literature results for these calibration fields is given in the Appendix (see Figures \ref{fig:A1}, \ref{fig:A2}, \ref{fig:A3}). For clarity, in Figure \ref{fig:A1} and  Figure \ref{fig:A2}, the cluster data are represented by a single point with the observed dispersion but our errors in the measured parameters have been estimated from the entire calibration data set.  The one-sigma errors for each parameter are indicated by the horizontal dashed lines. The metallicities of the globular clusters are adopted from the Harris database and the $\alpha$-enhancement is a median compilation of literature sources for each cluster. For the internal uncertainties, a temperature change of $\pm$ 100 K corresponds to a \logg\ change of $\pm$ 0.3 dex and a [Fe/H] change of $\pm$ 0.1.

\subsection[]{Linelist}

The adopted linelist comes from \cite{Kirby2008}, originally assembled from both the VALD \citet{Kupka1999} and the \citet{Kurucz1992} databases of atomic transitions. The VALD database was used for all neutral and singly ionized atoms and the \citet{Kurucz1992} database for CN, C$_2$, MgH as well as hyperfine transitions. The $\log gf$ values of the Fe lines and lines in surrounding regions were adjusted so that our synthetic spectra gave as good a match as possible to both the solar spectrum and the Arcturus spectrum (Hinkle Atlas) at our resolution of 11,000 and also to our spectra of the 16 bulge stars from Zoccali (see Appendix). The solar abundances were taken from MOOG which uses all of the \cite{Anders1989} values except for Fe.

At this resolution there is some blending of lines, predominantly with CN bands. The empirical $\log gf$ value adjustments are specific to this resolution and reflect the level of blending. The synthetic spectra are not a perfect match to the observed spectra for any particular set of stellar parameters; the $\log gf$ values were adjusted to give a good fit over the abundance range $-1.2 <$ [Fe/H] $< 0.1$ covered by the Sun, Arcturus and the Zoccali stars.  We then checked the stellar parameters returned by the $\chi^2$ program for our more metal-poor globular cluster stars at lower [Fe/H]. (We recall that the [Fe/H] values come only from Fe lines.) No obvious deviations were seen (see Figure \ref{fig:A2}). We conclude that the derived stellar parameters for giants are on a consistent abundance scale in the range $-2.1 <$ [Fe/H] $< 0.1$. Outside this abundance range, our stellar parameters are unconstrained by empirical calibration.

The derived stellar parameters from our $\chi^2$ program for the Sun are \logg\ $= 4.39 \pm 0.30$, [Fe/H] $= -0.02 \pm 0.09$, [$\alpha$/Fe] $= +0.04 \pm 0.10$ for a fixed \teff\ $= 5780$ K. For Arcturus the parameters are \logg\ $= 1.7 \pm 0.30$, [Fe/H] $= -0.5 \pm 0.09$, [$\alpha$/Fe] $= 0.20 \pm 0.10$ for \teff\ $= 4300$ K. These results compare well with the adopted parameters of \teff\ $= 5780$ K, \logg\ $= 4.44$, [Fe/H] $= 0.0$, [$\alpha$/Fe] =$ 0.0$ for the Sun and \teff\ $= 4286$ K, \logg\ $= 1.66$, [Fe/H] $= -0.52$ and [$\alpha$/Fe] $= 0.33$ for Arcturus \citep{Ramirez2011}. Note that the alpha enhancement values of Arcturus in the literature show some scatter. When this program was done in 2010, we were aiming to match Arcturus to the values of \citet{Peterson1993} of \teff\ $= 4270$ K, \logg\ $= 1.5$, [Fe/H] $= -0.5$ and [$\alpha$/Fe] $= 0.36$.  \citet{Norris1995} report an enhancement of 0.2 for the alpha elements of Ti, Si, Mg, closer to our derived value. \citet{Hill1997} similarly report lower enhancement values. We could not reproduce the higher value of \citet{Peterson1993} and similarly obtain a good match for our more alpha-enhanced open and globular cluster calibrators. Our linelist is available in the online manuscript and a sample of the list is shown in Table \ref{tab:Linelist}.

\subsection[]{Radial Velocity Measurement}

Radial velocities were measured from the sky-subtracted and wavelength-calibrated spectra using the IRAF program {\it fxcor}.  For the template spectra, a restricted grid of the synthetic spectra was used. The {\it fxcor} internal errors for our spectra are typically $0.5 - 0.7$ \kms, and the adopted template for each star was chosen to be the one for which the {\it fxcor} error was smallest. For fibre spectroscopy with a resolution of $11,000$ and mean S/N of $75$, we would expect total radial velocity errors of about $1$ \kms.  To estimate our total errors empirically, we compared the weighted mean velocity and velocity dispersion of the open cluster Melotte 66 measured by \citet{Sestito2008} (their radial velocities have errors of about $0.3$ \kms) with our measurement from $10$ stars in the cluster.  The mean difference in velocity (Sestito - ARGOS) is $-0.65 \pm 0.39$ \kms, and the ARGOS velocity error for a single measurement is $0.92 \pm 0.23$ \kms. 

\subsection[]{Determining Stellar Parameters}

Our $\chi^2$ procedure to derive the stellar parameters follows Kirby et al. (2008) who measured the metallicity of red giants from medium resolution spectra (R $\sim 8000$) over a wide metallicity range. This technique uses only weaker lines and was shown by Kirby et al. to reproduce the [Fe/H] measurements from high resolution spectra.  Our spectra show weak lines of Fe I, Ti I, Si I, Al I, Mg I, O I and CN and the three strong Ca-triplet lines. Specific regions of the spectra, excluding the strong lines have been selected over which to perform the $\chi^2$ minimization.  Excluding the strong lines and using only the Fe and weak line regions means we return a true [Fe/H] measurement. Additionally, we cannot simultaneously match our strong lines over the wide parameter space of our calibration stars in our synthetic spectra so we exclude these from the analysis \citep[see][]{Kirby2008}.

The procedure is iterative and requires at least four cycles to converge.  The temperature derived from $(J-K)_o$ depends on the current estimate of metallicity and gravity; the gravity, [Fe/H] and [$\alpha$/Fe] values derived from the $\chi^2$ process depend on the current temperature. The final cycle of the process gives the final adopted value of \logg\  which comes from the correct $\alpha$-enhanced grid. 

The $\alpha$-enhancement level significantly affects the stellar line profiles for the 
$\alpha$-elements, and it was essential to have grids of synthetic spectra that cover the full range of observed $\alpha$-enhancement from $0$ to $+1.0$.  

\begin{figure*}
   \centering
    \includegraphics[width=1.0\textwidth]{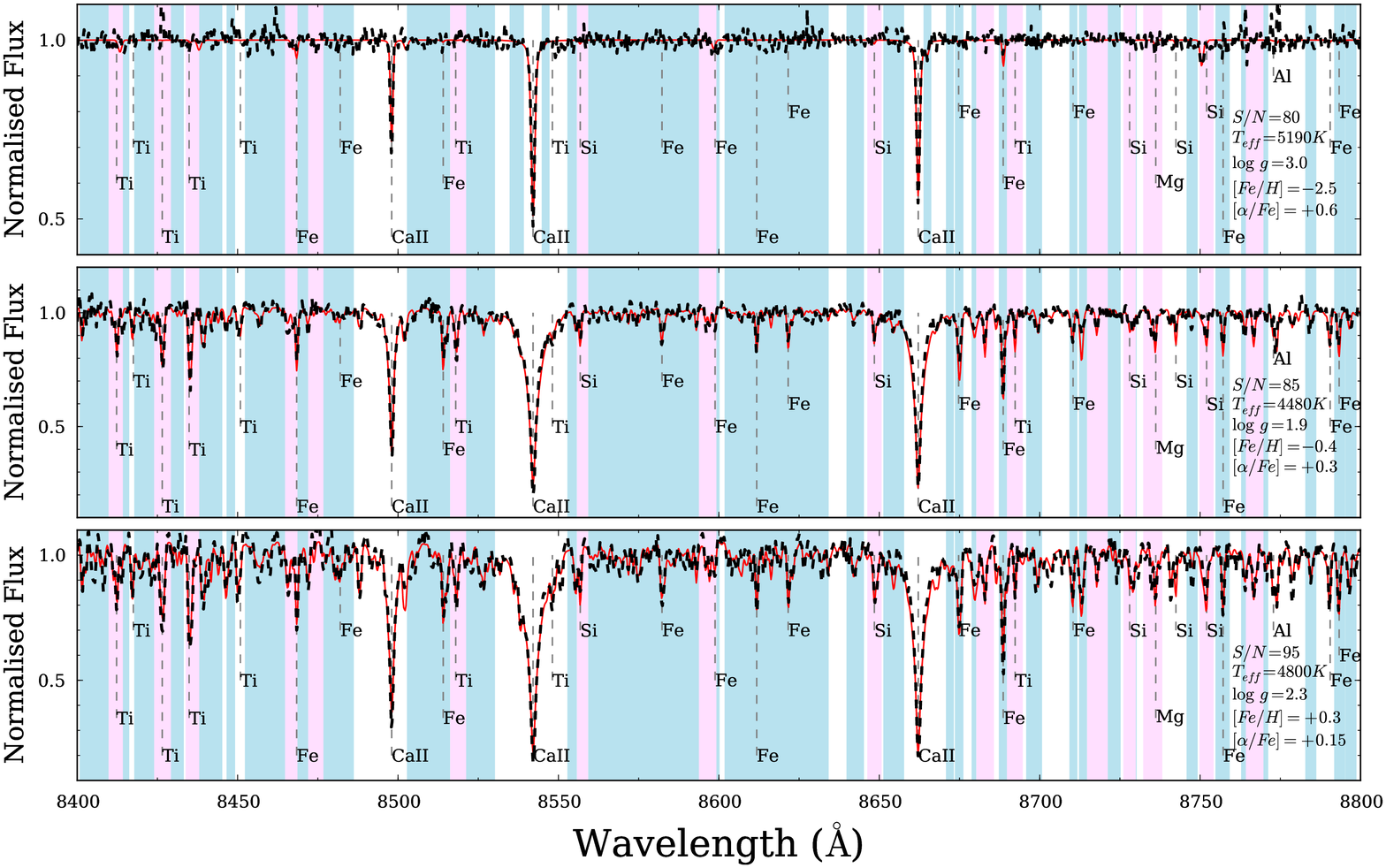}
 \caption{Ca-triplet region spectra for three stars in the sample. The star in the lower panel has \teff $= 4800$ K, 
\logg $= 2.3$, [Fe/H] $= 0.3$, $[\alpha$/Fe] $= 0.15$. The star in the intermediate panel has \teff $= 4480$ K, 
\logg $= 1.9$, [Fe/H] $= -0.4$, $[\alpha$/Fe] $= 0.3$ The star in the upper panel has  \teff $= 5190$ K, \logg $= 3.0$, [Fe/H] $= -2.5$, [$\alpha$/Fe] $= 0.6$. In each panel, the red spectrum is the synthetic spectrum for the adopted parameters and black dashed points is the spectrum as observed. The blue shaded regions correspond to the $\chi^2$ regions to determine [Fe/H] and the purple shaded regions correspond to the regions to determine the [$\alpha$/Fe]. Note the additional weak lines included near the Ca-triplet for stars $<$ --1.75, shown in the top panel. The most metal-poor star in the top panel represents the limit of [Fe/H] measurement using the weak line regions.} %
  \label{fig:twostars}
\end{figure*}

Figure \ref{fig:twostars} compares observed and synthetic spectra for three stars from our fields including two near the extremes of our [Fe/H] distribution. The star in the lower panel has [Fe/H] $= 0.3$ and [$\alpha$/Fe] $= 0.15$, the star in the intermediate panel has [Fe/H] $= -0.4$ and [$\alpha$/Fe] $= 0.3$ and the star in the top panel has [Fe/H] $\sim -2.5$ and a high alpha enhancement [$\alpha$/Fe] $= 0.6$. At this resolution and with a S/N $= 50$, the [Fe/H] value for the metal-poor star in the upper panel of Figure \ref{fig:twostars} is at the lower limit of measurability. For this star, with \teff\ $\sim 5190$ K, the Fe lines are barely visible against the noise, and the program cannot differentiate abundances lower than [Fe/H] $= -2.5$.  For a S/N of $80$, [Fe/H] values of $\sim -2.8$ can still be measured. Stars of yet lower metallicity can be identified in our survey but their metallicities will be uncertain.

We tested the dependence of the [$\alpha$/Fe] ratio on S/N using synthetic spectra with noise added to the spectra and run through the $\chi^2$ process.  For stars with S/N $= 50$ per resolution element and [Fe/H] $< -2.0$, the uncertainty of the $\alpha$-enhancement above our nominal reported error in [$\alpha$/Fe] of 0.10 is increased by about $0.1$ dex. For S/N $= 80$, it is increased by about $0.05$. For stars with [Fe/H] $> -1.0$, there is no dependence of uncertainty on S/N for spectra with S/N $> 50$.

\noindent{The steps in the $\chi^2$ process are: }\\
\noindent 1a) Choose wavelength regions for calculating $\chi^2$ for determination of [Fe/H] (weak Fe and weakly blended lines only). Exclude the Ca-triplet and lines of Ti, Al, Mg, Si. The regions used in the determination of [Fe/H] are shown in Figure  \ref{fig:twostars}. There are about 30 Fe lines across this region. The equivalent widths of these lines is up to 500 m\AA\ for stars around solar metallicity and decreases to 200 m\AA\ for stars with [Fe/H] = --1.0. \\
1b) Choose wavelength regions over which to calculate $\chi^2$ for measurement of [$\alpha$/Fe] using a combination of the 17 available weak Ti, Si, Mg lines in the synthetic spectra. These regions are shown in Figure  \ref{fig:twostars}. The equivalent width of the alpha-element lines is up to 500 m\AA\ at solar metallicity and about 200 m\AA\ at [Fe/H] = --1.0, similar to iron. \\
2) Shift all observed spectra back to rest-frame wavelengths. \\
3) Mask out any residual sky lines. \\
4) Interpolate synthetic spectra on to the wavelength grid of the observed spectrum ($0.24$\AA\ channels) and measure resolution of observed spectra (R is nominally $\sim 11,000$ but varies by up to $10$\%). Then convolve synthetic spectra to resolution of the observed spectrum. \\
5) Normalise observed and synthetic spectra in the same way using continuum regions identified from \cite{Hinkle2005} for Arcturus and the Sun. \\
6) Make a first estimate of the temperature from the $(J - K)_o$ colour. The interstellar reddening for each star was determined from the \citet{Schlegel} reddening maps. The relationship between colour and temperature was taken from the \cite{Bessell1998} calibration, which was calculated using the Castelli/Kurucz model atmospheres. Their calibration was transformed from the \citet{Bessell1998} $(J-K)$ system to the 2MASS system using the tranformation $(J-K)_{\rm 2MASS} = 0.983 (J-K)_{\rm BCP}$ \citep{Carpenter2001}. The \teff\ calibration depends on [Fe/H] and \logg, and values of [Fe/H] $= -0.5$ and \logg\ $= 2.5$ were used to start the iteration. \\
7) Compare each star with the synthetic spectra in the $[\alpha/Fe]= 0.0 $ grid and return the $\chi^2$ value over the [Fe/H] regions from each comparison: 

$$\chi^2 = \sum_{i=1}^k \left(\frac{X_i - O_i}{\sigma_i} \right)^2$$

\noindent where $i = 1 ... k$ are the wavelength channels for the [Fe/H] regions, $X_i$ is the synthetic spectrum, $O_i$ is the observed spectrum and $\sigma_i$ is the standard deviation of the observed spectra for a Poisson-noise dominated system.\\
8) Interpolate to find the minimum value of $\chi^2$ in the two dimensions of \logg\ and [Fe/H] to derive the best fit stellar parameters. \\
9) With the new [Fe/H] value for the star, redetermine \teff\ using the \cite{Bessell1998} calibration.  This second estimate is the final temperature to be used, although the gravity will change slightly when the final 
\logg\ value is determined from the $\chi^2$ comparison of the star with the appropriate $\alpha$-enhanced grid. \\
10) Do a second $\chi^2$ iteration over a narrower stellar parameter range of the synthetic grid, around the minima of the first pass results, and interpolate over [Fe/H], $\log g$ and [$\alpha$/Fe], using the Fe regions for the [Fe/H] estimate and the $\alpha$-regions of Ti, Si and Mg for [$\alpha$/Fe].\\
11). With this [$\alpha$/Fe] value, use the appropriate $\alpha$-enhanced grid to derive the final $\log g$ value.

The procedure in step 11 was done in two independent ways which agree within $\pm$ 0.25 dex in \logg. The first implementation used the correct [$\alpha$/Fe] grid and the whole spectral region to return the \logg\ value.  The reason for using this method was to make use of the gravity information in the wings of the strong Ca-triplet lines. For this step, the \logg$f$ values for the triplet lines (included in the linelist) were adjusted so that the $\chi^2$ program returned the correct $\log g$ values for the Sun and Arcturus for an [$\alpha$/Fe] grid of 0 and 0.2, respectively. The [M/H] returned from the $\chi^2$ comparison to obtain the $\log g$ using the entire spectral region is disregarded as the $\log g$ and [M/H] could not be simultaneously well determined over the wide parameter space of the calibration set.

The second method fixed the [Fe/H] for the star and then interpolated only the $\chi^2$ iron and $\alpha$-element regions shown in Figure  \ref{fig:twostars} to determine the minimum $\chi^2$ for the single parameter \logg. These weak line features are not as strongly gravity--sensitive as the Ca-triplet. However interpolation in this single dimension enabled the \logg\ value to be returned again to within $\pm$ 0.25 of the values of the first method. 

\noindent{12) As described in section 4.1,  the $\chi^2$ program was checked against spectra of stars with known parameters over the abundance range [Fe/H] $=-2.1$ to $0.1$.  See the Appendix for comparison of the $\chi^2$ and literature results for [Fe/H], [$\alpha$/Fe] and \logg\ for the set of calibration stars.} 

\section{Elimination of Foreground Dwarfs}

The low resolution spectra ($R \sim 3000$) from 5000 -- 5600\AA\ were used for a first-pass elimination of foreground G and K dwarfs from the sample by computing an index that measures the strength of the gravity-sensitive Mgb/MgH feature. The index was based on synthetic spectra of dwarf stars ($\log g$ $\sim$ 4) and typical clump stars ($\log g$ $\sim$ 2.7). However, this was only partly successful. Due the large range in stellar parameters of stars in the sample, it was not possible to cleanly discriminate against the dwarfs. In order to ensure that the more metal-rich ([Fe/H] $> 0$) giant stars were not excluded from the sample, it was necessary to set the index cut such that the more metal-poor dwarfs ([Fe/H] $< -0.5$) would also pass through. Additionally, for about $30$\% of the stars, the signal level in the low resolution blue band  was not high enough to allow secure discrimination of dwarfs and giants. The $\chi^2$ program applied to the Ca-triplet region spectra was, however, easily able to identify the residual dwarfs in the sample.

\section{Stellar Distances}

Our selection criteria were aimed primarily at selecting red clump stars, because for these stars we can derive relatively accurate photometric distances. The sample does, however, include some fainter and brighter giants and also some foreground dwarfs.  We use our estimates of stellar parameters \teff, \logg, [Fe/H] and [$\alpha$/Fe] to identify the red clump stars.   The study by \citet{Alves2000} of a sample of Hipparcos red clump stars with $-0.6 <$ [Fe/H] $< 0.1$ gave a mean absolute K-magnitude of $M_K = -1.61 \pm 0.03$, and we adopt this mean value independent of metallicity \citep[see also][]{McWilliam2010}. Alves fitted a gaussian to the luminosity function of his stars and found a dispersion in $M_K$ of $0.22$ mag.  The mean error in the parallaxes is about 5\%, which contributes about $0.11$ mag to the observed dispersion of the red clump $M_K$ values.  For estimating the uncertainty in the distances of our clump stars, we adopt Alves's intrinsic dispersion of $0.22$ mag for the $M_K$ values.  This is similar to dispersions adopted by \citet{Stanek1997} and \citet{Babusiaux2005}.  For the clump stars in our sample, the intrinsic dispersion of $0.22$ in $M_K$ is the dominant error source for the distance estimates.

The mean \logg\ for the red clump depends weakly on age and metallicity but is typically about 2.4. We can estimate the expected intrinsic spread in \logg\ from the dispersion in $M_K$, using the \citet{Marigo2008} isochrones for [Fe/H] = --0.2 and age 10 Gyr.  Along the entire giant branch, the isochrones show that $M_K$ increases almost linearly with \logg, with 
d$M_K$/d(\logg) = 2.3.  The dispersion of $0.22$ mag in $M_K$ corresponds to a dispersion in \logg\ of about $0.1$. In the (\logg, \teff) plane, the He-core burning red clump stars are superimposed on a background of first ascent giants. This background of first giants can be minimised by choosing an appropriately narrow interval of \logg:  a fairly clean sample of red clump stars could be selected by taking stars within say $\pm 0.2$ dex of the mean \logg\ value, if the \logg\ values were accurate enough to justify such a choice.  Our measuring errors in \logg\ are about $0.30$ mag (see Appendix), so we take stars with \logg\ $=1.8$ to $3.2$ and \teff\ $=4500$ to $5300$ K as potential clump giants.

\begin{figure}
\centering
\includegraphics[width=0.45\textwidth]{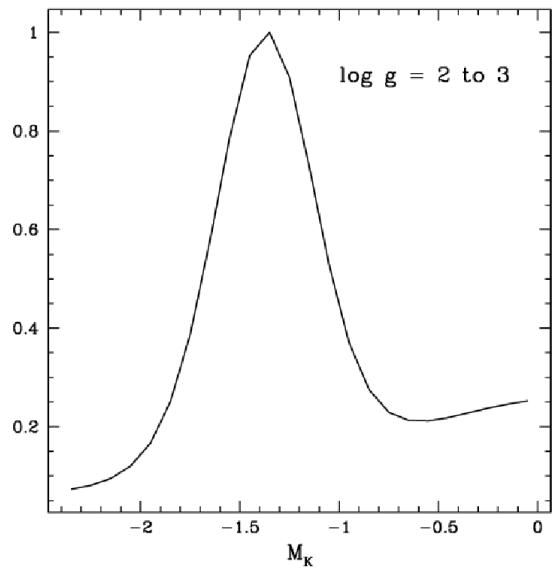} 
\caption{The expected distribution of $M_K$ for an unbiased populations of stars with [Fe/H] $= -0.2$ and age 10 Gyr lying on the giant branch with  \logg\ $=2$ to $3$ (derived from the \citet{Marigo2008} isochrones). The luminosity function for the red clump stars has been convolved with a gaussian with $\sigma = 0.22$ as for the Alves (2000) study.  The mean $M_K$ for the isochrone is slightly displaced from the \citet{Alves2000} value.}
 \label{fig:prob}
\end{figure}

In the above interval of \logg\ and \teff, we have no way of telling which star is a red clump star and which is a first ascent giant.  Near the mean \logg\ of the red clump stars, the first ascent giants have almost the same value of $M_K$ as the red clump stars themselves, but this is not so for first ascent giants with \logg\ values towards the limits of our interval.  As a guide to the size of the effect, we can use the \citet{Marigo2008} isochrones to estimate the probability distribution of $M_K$ for the stars that lie on the giant branch and have $2 <$ \logg\ $< 3$. This is shown in Figure \ref{fig:prob} for a unbiased population with [Fe/H] = $-0.2$ and age 10 Gyr.   Our sample is biased against the fainter stars because of the incompleteness of the 2MASS survey at the fainter apparent magnitudes, so the numbers of intrinsically fainter giants with $2 <$ \logg\ $< 3$ in our sample is reduced relative to the distribution shown in Figure \ref{fig:prob}.

For the purposes of this survey, we take all of the potential red clump stars to have $M_K = -1.61 \pm 0.22$, recognising that for some of the underlying first ascent giants the uncertainty will be somewhat larger.

For stars that are not on the red clump, we derive $M_K$ values using BaSTI isochrones of age $10$ Gyr and appropriate [Fe/H].  The stars lie fairly close to the isochrones in the (\teff,\,\logg) plane (see paper III), and their absolute magnitudes are derived by adjusting the temperature of the star to place it on the closest branch of the isochrone.   The uncertainty in \logg\ is $0.30$ (see section 4.1) so the uncertainty in distance for non-clump giants is about $0.70$ mag.

In summary, for the clump star candidates which comprise about $70$\% of our sample, the adopted distance error is $\sigma_K = 0.22$ mag (about 900 pc at the distance of the Galactic centre). For the non-clump giants, the distance error is $\sigma_K = 0.70$ mag, dominated by the uncertainty in the \logg\ estimates. The near-linear relationship between $M_K$ and \logg\ on the giant branch means that the error distributions for the distance moduli for both clump and non-clump giants are close to Gaussian.
 
\section{Conclusion}

In this paper we have presented the ARGOS survey. Papers III (metallicity components) and IV (kinematics) will show that we have achieved our goals of finding  the red clump giants in the bulge and discovering the small fraction of metal-poor stars in the bulge region: $\sim$ 0.15$\%$ of our stars have [Fe/H] $< -2.0$. These later papers will describe the components of the metallicity distribution function that are seen throughout the bulge region, discuss their origin in the context of bulge formation theory, and show how the kinematics of the bulge stars change with metallicity. 

\section*{Acknowledgements}

We thank the Anglo-Australian Observatory, who have made this project possible.

This publication makes use of data products from the Two Micron All Sky Survey, which is a joint project of the University of Massachusetts and the Infrared Processing and Analysis Center/California Institute of Technology, funded by the National Aeronautics and Space Administration and the National Science Foundation.

This work has been supported by the RSAA and Australian Research Council grant DP0988751. 
EA gratefully acknowledges financial support by the European Commission through the DAGAL Network (PITN-GA-2011-289313). J.BH is supported by an ARC Federation Fellowship. M.A gratefully acknowledges support
through an ARC Laureate Fellowship. G.F.L thanks the Australian research council for support through his Future Fellowship (FT100100268) and Discovery Project (DP110100678). R.R.L gratefully acknowledges support from the Chilean {\sl Centro de Astrof\'\i sica} FONDAP No. 15010003. L.L.K is supported by the Lend\"ulet program of the Hungarian Academy of Sciences and the Hungarian OTKA Grants K76816, MB08C 81013 and K83790. Discussions with Mary Williams about red clump giants are gratefully acknowledged. We thank the referee for comments that
improved the presentation of the paper.

\appendix
\section{Appendix}

The uncertainties in our estimates of [Fe/H], [$\alpha$/Fe] and \logg\ were determined by comparing our parameter estimates with those in the literature for the Sun, Arcturus, 16 bulge stars observed by Zoccali et al (2008), the open cluster Melotte 66 and six globular clusters. Figures A1--A3 show the differences $\Delta$ between our $\chi^2$ estimates and the literature estimates for [Fe/H] and [$\alpha$/Fe], and between our $\chi^2$ and comparison estimates for \logg.  For the [$\alpha$/Fe] ratio we use the mean of the Mg, Si and Ti abundances. 

The scatter of the differences $\Delta$ comes from our $\chi^2$ measuring errors and errors in the comparison values of the parameters
$$ \sigma^{2}_\Delta = \sigma^{2}_{\chi^2} + \sigma{^2}_{comp} $$
\noindent where $\sigma_\Delta$ is the weighted dispersion of the differences $\Delta$.
For [Fe/H], $\sigma_\Delta = 0.13$ and we estimate that $\sigma_{comp} = 0.10$, so $\sigma_{\chi^2} = 0.09$.  For [$\alpha$/Fe], we find that $\sigma_{\chi^2} = 0.10$.  

For the \logg\ errors, some of our comparison stars are in clusters and we compared our $\chi^2$ estimates with \logg\ values from isochrone fits using the \citet{Marigo2008} isochrones.  Colour
and magnitude data and the adopted cluster ages and distances were taken from the literature. 
For example, the age of Melotte 66 was taken to be $4 \pm 1$ Gyr and its distance modulus $(m-M)_o = 13.2~(+0.3,-0.1$) \citep{Kassis1997}. The stars fit well on the isochrone after small adjustments within the errors were made to the distance. The main source of error in the \logg\ determination for this procedure is the uncertainty in the age of the population. For Melotte 66, an age uncertainty of $\pm 1$ Gyr corresponds to an uncertainty $\sigma_{\log g} = 0.20$ dex. For the older clusters, an age error of $1$ Gyr corresponds to a lower uncertainty $\sigma_{\log g} = 0.13$. From the distribution of differences shown in Figure A3, we estimate that our uncertainty on \logg\ from the $\chi^2$ process is $\sigma_{\log g} = 0.30$ dex.

\begin{figure*}
\centering 
\vspace{-10pt}
\includegraphics[width=0.7\textwidth]{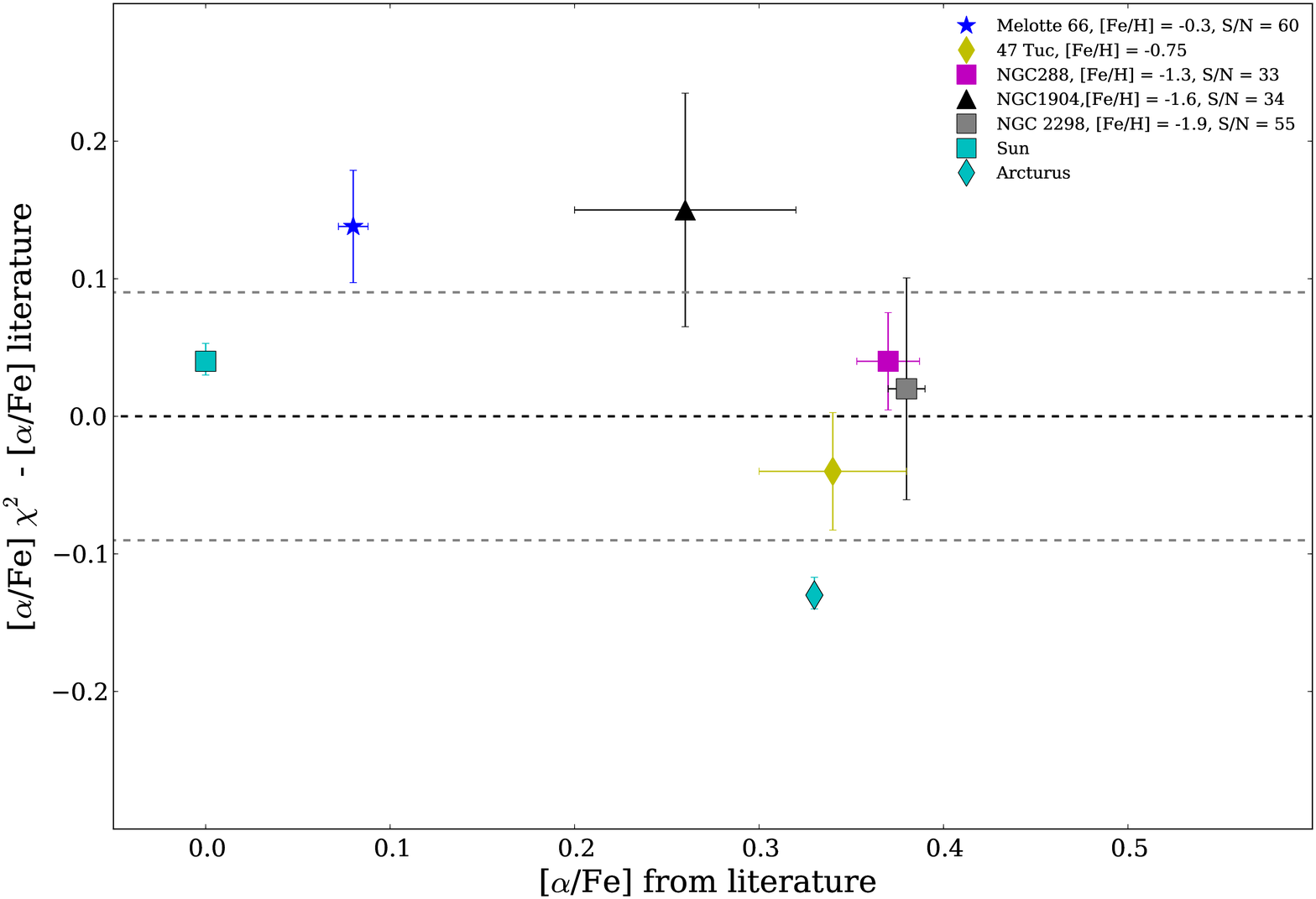}
\caption{[$\alpha$/Fe] calibration. For the open and globular cluster data, the error bars in the $x$-direction represent the measurement error of the literature comparison values $\sigma_{comp}$, and the error bars in the $y$-direction represent the combined internal $\chi^2$ error  and the measuring errors for the comparison values, $\sigma_{i}^2$ = ($\sigma_{comp}^2$ + $\sigma_{\chi^2}^2$). For Arcturus and the Sun, the errors in the $y$-direction are the internal $\chi^2$ fitting errors.  }
\label{fig:A1}
\end{figure*}
     
\begin{figure*}
\centering
 \vspace{-10pt}
\includegraphics[width=0.7\textwidth]{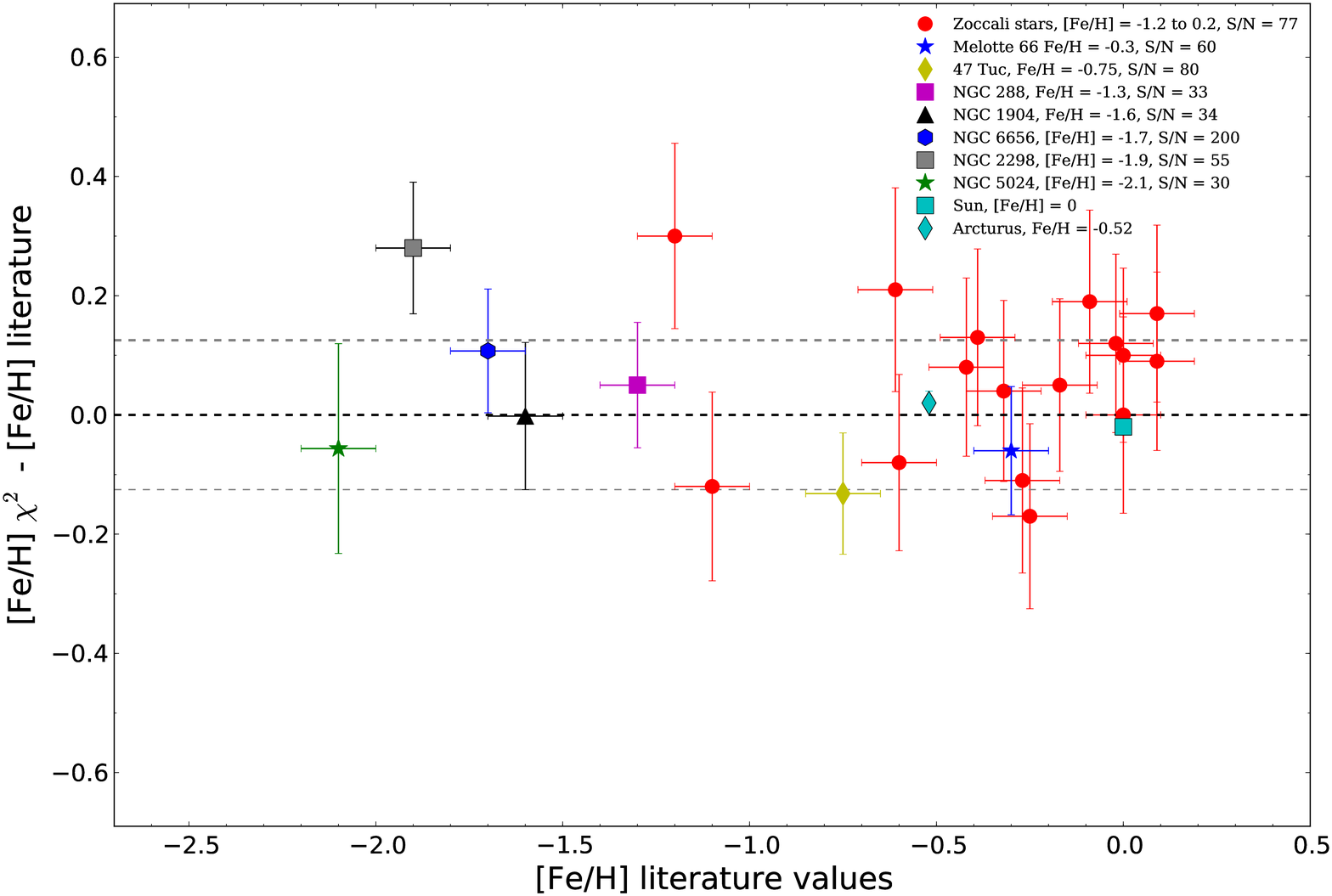}
\caption{[Fe/H] calibration. The error bars in the $x$-direction represent the estimated literature errors. As for Figure A1, the error bars in the direction of the $y$-axis represent the combined literature error and the $\chi^2$ fitting errors for the individual stars and the measuring error for the open and globular clusters, $\sigma_{\chi^2}$, of their mean [Fe/H].}
\label{fig:A2}
\end{figure*}

\begin{figure*}
\centering
\vspace{-10pt}      
\includegraphics[width=0.7\textwidth]{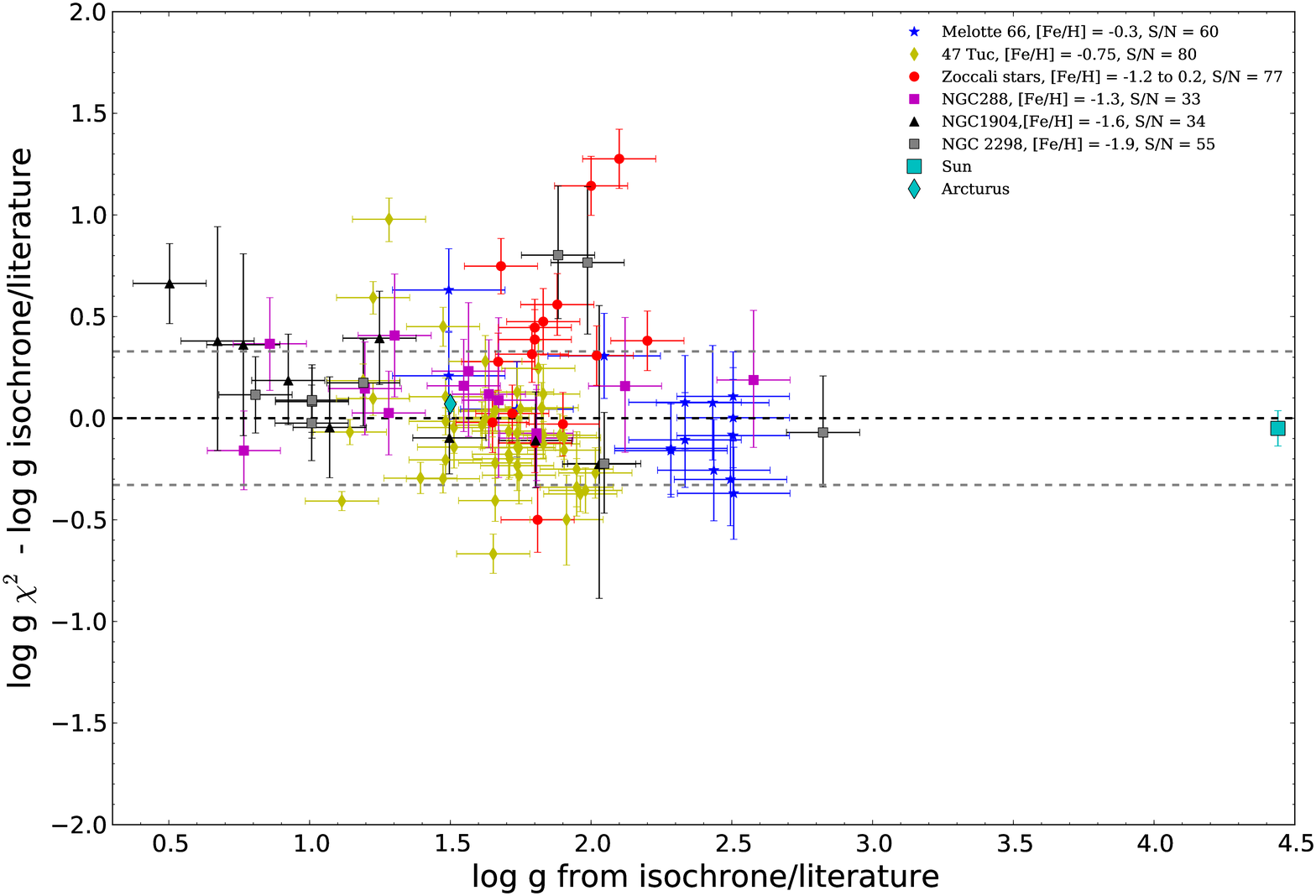}
\caption[]{\logg\ calibration. The error bars in the $x$-direction represent the uncertainty in the estimated \logg\ for the calibration stars. The error bars in the $y$-direction represent the combined error of the $\chi^2$ program and the uncertainty of the \logg\ value for the comparison objects.  Note the atypically large errors from the poor S/N spectra for the cluster NGC1904. The \logg\ values for the Sun and Arcturus and \citet{Zoccali2008} are taken from the literature and all \logg\ values for cluster stars are from isochrones estimates (see text)$^3$.} 

\label{fig:A3}
\end{figure*}
\footnotetext[3]{There are two \citet{Zoccali2008} stars that lie clearly above the $1-\sigma$ errorbars of the distribution in the figure. One of these stars falls on the RGB branch from the ARGOS calibration and the AGB branch for the \citet{Zoccali2008} calibration. The other outlier is a poor match to the isochrone for both ARGOS and \citet{Zoccali2008} calibrations.}
     
\begin{table*}
 \centering
 \caption{Comparison of stellar parameters for the 16 stars observed for ARGOS measured by \citet{Zoccali2008}.}
    \begin{tabular}{| c | c | c | c | c| c| c | c | c| }
  \hline
  \multicolumn{2}{|c|}{} &  \multicolumn{4}{|c|}{\citet{Zoccali2008}} & \multicolumn{3}{|c|}{ARGOS}\\
  \hline
 OGLE & S/N &  \teff\ & \logg\ & [Fe/H] & \teff & \logg\ & [Fe/H] & [$\alpha$/Fe] \\
 \hline
100047c6 & 80 & 4350 & 1.72 & -0.33 & 4470 & 1.7 & -0.26 & 0.3\\
104943c6 & 75 & 5300 &  2.1 & -1.2 & 5660 & 3.4 & -0.90 & 0.2\\
75097c7 & 65 & 4500 &  1.8 & -0.09 & 4650  & 2.2 & 0.10& 0.1\\
57883c7 & 65 & 4550 & 1.8 & -0.25 & 4440 & 1.6 & -0.44 &  0.3 \\
90337c7 & 70 & 4850 & 2.02 & -0.27 & 4770  & 2.3& -0.36 & 0.2\\
64860c7 & 75  & 4500 & 1.8 & 0.09 & 4790 & 2.2  & 0.26 & 0.2 \\
98458c7 & 80 & 4300 &  1.68 & -0.02 & 4540 & 2.4 & 0.1 & 0.1\\
87242c8 & 70 & 4300 & 1.67 & 0 & 4440 & 2.0 & 0.1 &  0.2\\
43791c3 & 65 & 4950 & 2.2 & 0 & 5160 & 2.6 & 0.0 &  0.1\\
32080c3 & 85 & 5000 & 2.0 & -0.17 & 5450 & 3.1  & -0.12 &  0.2\\
23017c3 & 80 & 4250 & 1.65 & 0.09 & 4410  & 1.6 & 0.18 & 0.1\\
38354c3 & 75 & 4700 & 1.83 & -0.61 & 4620  & 1.9 & -0.68 &  0.2\\
76187c3 & 80 & 4550 &  1.79 & -0.42 & 4700  & 2.1 & -0.34 &  0.4\\
56410c2 & 75 & 4600 & 1.81 & -1.1 & 4570 & 1.3 & -1.20 &  0.5\\
19402c1 & 60 & 4550 & 1.83 & -0.61 & 4730  & 2.3 & -0.40 & 0.2\\
212324c6 & 75 & 4800 & 1.88 & -0.32 & 4960  & 2.4 & -0.28 & 0.2\\                                                    
\hline
\end{tabular}
\centering
\label{table:Zoc}
\end{table*}     

\begin{table*}
 \centering
\caption{Sample of the linelist available in full in the electronic version of this paper. Following \citet{Kirby2008}, changes from original log $gf$ values are indicated by values with single decimal places. The linelist is in MOOG readable format.    }
    \begin{tabular}{| c | c | c | c | }
  \hline
Wavelength ($\AA$)& Species & EP (eV) & log $gf$  \\
 \hline
  8467.937   & 20.000   &  4.554  &  -3.313 \\
  8468.407  &  26.000   &  2.223   & -2.3 \\
  8468.470   & 22.000   &  1.887   & -1.2 \\
  8468.592  &  26.100   &  9.700  &  -3.0 \\
  8468.657  &  607.00   &  0.916  &  -1.597 \\
  8468.732  &  28.100   &  6.952  &  -4.8 \\
  8468.832   & 25.000   &  6.412  &  -2.8 \\
  8468.849   & 12.000   &  5.932  &  -2.240 \\
  8468.872   & 607.00  &   0.916  &  -2.9 \\
\hline
\end{tabular}
\centering
\label{tab:Linelist}
\end{table*}


\begin{thebibliography}{99}


\bibitem[\protect\citeauthoryear{Alonso, Arribas, 
\& Mart{\'{\i}}nez-Roger}{1999}]{Alonso1999} Alonso A., Arribas S., Mart{\'{\i}}nez-Roger C., 1999, A\&AS, 140, 261 

\bibitem[\protect\citeauthoryear{Alves}{2000}]{Alves2000} Alves 
D.~R., 2000, ApJ, 539, 732 

\bibitem[\protect\citeauthoryear{Alves-Brito et 
al.}{2010}]{AlvesBrito2010} Alves-Brito A., Mel{\'e}ndez J., Asplund M., Ram{\'{\i}}rez I., Yong D., 2010, A\&A, 513, A35 


\bibitem[\protect\citeauthoryear{Anders 
\& Grevesse}{1989}]{Anders1989} Anders E., Grevesse N., 1989, GeCoA, 53, 197 

\bibitem[\protect\citeauthoryear{Athanassoula}{2005}]{Athanassoula2005} 
Athanassoula E., 2005, MNRAS, 358, 1477 


\bibitem[\protect\citeauthoryear{Babusiaux 
\& Gilmore}{2005}]{Babusiaux2005} Babusiaux C., Gilmore G., 2005, astro, arXiv:astro-ph/0506413 


\bibitem[\protect\citeauthoryear{Babusiaux et al.}{2010}]{Babusiaux2010} Babusiaux C., et al., 2010, A\&A, 519, A77 

\bibitem[\protect\citeauthoryear{Bensby et 
al.}{2010}]{Bensby2010} Bensby T., Alves-Brito A., Oey M.~S., Yong D., Mel{\'e}ndez J., 2010, A\&A, 516, L13 


\bibitem[\protect\citeauthoryear{Bessell 
\& Brett}{1988}]{BessellBrett1988} Bessell M.~S., Brett J.~M., 1988, PASP, 100, 1134 

\bibitem[\protect\citeauthoryear{Bessell, Castelli, 
\& Plez}{Bessell et~al.}{1998}]{Bessell1998} Bessell M.~S., Castelli F., Plez B., 1998, wwwuser.oat.ts.astro.it/castelli/colours/bcp.html


 \bibitem[\protect\citeauthoryear{Brook et al.}{2007}]{Brook2007} 
Brook C.~B., Kawata D., Scannapieco E., Martel H., Gibson B.~K., 2007, ApJ, 
661, 10 

\bibitem[\protect\citeauthoryear{Carpenter}{2001}]{Carpenter2001} 
Carpenter J.~M., 2001, AJ, 121, 2851 

\bibitem[\protect\citeauthoryear{Castelli 
\& Cacciari}{2001}]{Castelli2001} Castelli F., Cacciari C., 2001, A\&A, 380, 630 

\bibitem[\protect\citeauthoryear{Castelli 
\& Kurucz}{2004}]{Castelli2004} Castelli F., Kurucz R.~L., 2004, astro, arXiv:astro--ph/0405087 

\bibitem[\protect\citeauthoryear{Diemand, Madau, 
\& Moore}{2005}]{Diemand2005} Diemand J., Madau P., Moore B., 2005, MNRAS, 364, 367 

\bibitem[\protect\citeauthoryear{Fulbright, McWilliam, 
\& Rich}{2007}]{Fulbright2007} Fulbright J.~P., McWilliam A., Rich R.~M., 2007, ApJ, 661, 1152 

\bibitem[\protect\citeauthoryear{Gonzalez et 
al.}{2011}]{Gonzalez2011} Gonzalez O.~A., Rejkuba M., Zoccali M., Valenti E., Minniti D., 2011, A\&A, 534, A3


\bibitem[\protect\citeauthoryear{Gratton 
\& Ortolani}{1989}]{Gratton1989} Gratton R.~G., Ortolani S., 1989, A\&A, 211, 41 

\bibitem[\protect\citeauthoryear{Gratton 
\& Contarini}{1994}]{Gratton1994} Gratton R.~G., Contarini G., 1994, A\&A, 283, 911 

\bibitem[\protect\citeauthoryear{Haywood}{2008}]{Haywood2008} 
Haywood M., 2008, MNRAS, 388, 1175 

\bibitem[\protect\citeauthoryear{Hill}{1997}]{Hill1997} Hill V., 1997, A\&A, 324, 435 

\bibitem[\protect\citeauthoryear{Hinkle et al.}{2005}]{Hinkle2005} 
Hinkle K., Wallace L., Valenti J., Ayres T., 2005, uaas.book,  

\bibitem[\protect\citeauthoryear{Howard et al.}{2009}]{Howard2009} 
Howard C.~D., et al., 2009, ApJ, 702, L153 

\bibitem[\protect\citeauthoryear{Johnson et 
al.}{2011}]{Johnson2011} Johnson C.~I., Rich R.~M., Fulbright 
J.~P., Valenti E., McWilliam A., 2011, ApJ, 732, 108 

\bibitem[\protect\citeauthoryear{Kassis et al.}{1997}]{Kassis1997} 
Kassis M., Janes K.~A., Friel E.~D., Phelps R.~L., 1997, AJ, 113, 1723 

\bibitem[\protect\citeauthoryear{Kirby, Guhathakurta, 
\& Sneden}{2008}]{Kirby2008} Kirby E.~N., Guhathakurta P., Sneden C., 2008, ApJ, 682, 1217 

\bibitem[\protect\citeauthoryear{Kurucz}{1992}]{Kurucz1992} Kurucz 
R.~L., 1992, RMxAA, 23, 45 

\bibitem[\protect\citeauthoryear{Kupka 
\& Ryabchikova}{1999}]{Kupka1999} Kupka F., Ryabchikova T.~A., 1999, POBeo, 65, 223 

\bibitem[\protect\citeauthoryear{Lane et 
al.}{2011}]{Lane2011} Lane R.~R., Kiss L.~L., Lewis G.~F., Ibata R.~A., Siebert A., Bedding T.~R., Sz{\'e}kely P., Szab{\'o} G.~M., 2011, A\&A, 530, A31 

\bibitem[\protect\citeauthoryear{Marigo et 
al.}{2008}]{Marigo2008} Marigo P., Girardi L., Bressan A., Groenewegen M.~A.~T., Silva L., Granato G.~L., 2008, A\&A, 482, 883 

\bibitem[\protect\citeauthoryear{Martinez-Valpuesta, Shlosman, 
\& Heller}{2006}]{Inma2006} Martinez-Valpuesta I., Shlosman I., Heller C., 2006, ApJ, 637, 214 

\bibitem[\protect\citeauthoryear{Martinez--Valpuesta 
\& Gerhard}{2011}]{Inma2011} Martinez--Valpuesta I., Gerhard O., 2011, ApJ, 734, L20 


\bibitem[\protect\citeauthoryear{McWilliam, Geisler, 
\& Rich}{1992}]{McWilliam1992} McWilliam A., Geisler D., Rich R.~M., 1992, PASP, 104, 1193 

\bibitem[\protect\citeauthoryear{McWilliam 
\& Bernstein}{2008}]{McWilliam2008} McWilliam A., Bernstein R.~A., 2008, ApJ, 684, 326 


\bibitem[\protect\citeauthoryear{McWilliam 
\& Zoccali}{2010}]{McWilliam2010} McWilliam A., Zoccali M., 2010, ApJ, 724, 1491 


\bibitem[\protect\citeauthoryear{Mel{\'e}ndez et 
al.}{2008}]{Melendez2008} Mel{\'e}ndez J., et al., 2008, A\&A, 484, L21 

\bibitem[\protect\citeauthoryear{Minniti et 
al.}{1995}]{Minniti1995} Minniti D., Olszewski E.~W., Liebert J., 
White S.~D.~M., Hill J.~M., Irwin M.~J., 1995, MNRAS, 277, 1293 

\bibitem[\protect\citeauthoryear{Nataf et al.}{2010}]{Nataf2010} 
Nataf D.~M., Udalski A., Gould A., Fouqu{\'e} P., Stanek K.~Z., 2010, ApJ, 
721, L28 

\bibitem[\protect\citeauthoryear{Ness et al.}{2012}]{Ness2012} 
Ness M., et al., 2012a, ApJ, 756, 22 


\bibitem[\protect\citeauthoryear{Norris 
\& Da Costa}{1995}]{Norris1995} Norris J.~E., Da Costa G.~S., 1995, ApJ, 447, 680 


\bibitem[\protect\citeauthoryear{Peterson, Dalle Ore, 
\& Kurucz}{1993}]{Peterson1993} Peterson R.~C., Dalle Ore C.~M., Kurucz R.~L., 1993, ApJ, 404, 333 

\bibitem[\protect\citeauthoryear{Ram{\'{\i}}rez 
\& Allende Prieto}{2011}]{Ramirez2011} Ram{\'{\i}}rez I., Allende Prieto C., 2011, ApJ, 743, 135 


\bibitem[\protect\citeauthoryear{Saha, Martinez-Valpuesta, 
\& Gerhard}{2012}]{Saha2012} Saha K., Martinez-Valpuesta I., Gerhard O., 2012, MNRAS, 421, 333 


\bibitem[\protect\citeauthoryear{Saito et al.}{2011}]{Saito2011} 
Saito R.~K., Zoccali M., McWilliam A., Minniti D., Gonzalez O.~A., Hill V., 
2011, AJ, 142, 76


\bibitem[\protect\citeauthoryear{{Schlegel}, {Finkbeiner}, 
\& {Davis}}{{Schlegel} et~al.}{1998}]{Schlegel} Schlegel D.~J., Finkbeiner D.~P., Davis M., 1998, ApJ, 500, 525 

\bibitem[\protect\citeauthoryear{Sestito et al.}{2008}]{Sestito2008}Sestito, P.,  Bragaglia, A., Randich, S., Pallavicini, R., Andrievsky, S.M., Korotin, S. A., 2008, A\&A, 488, 943


\bibitem[\protect\citeauthoryear{Sharma et al.}{2011}]{Sharma2011} 
Sharma S., Bland-Hawthorn J., Johnston K.~V., Binney J., 2011, ApJ, 730, 3 

\bibitem[\protect\citeauthoryear{Sharp et al.}{2006}]{Sharp2006} 
Sharp R., et al., 2006, SPIE, 6269,  


\bibitem[\protect\citeauthoryear{Shen et al.}{2010}]{Shen2010} 
Shen J., Rich R.~M., Kormendy J., Howard C.~D., De Propris R., Kunder A., 
2010, ApJ, 720, L72 

\bibitem[\protect\citeauthoryear{Shetrone 
\& Keane}{2000}]{Shetrone2000} Shetrone M.~D., Keane M.~J., 2000, AJ, 119, 840 


\bibitem[\protect\citeauthoryear{Skrutskie et 
al.}{2006}]{2MASS} Skrutskie M.~F., et al., 2006, AJ, 131, 
1163 

\bibitem[\protect\citeauthoryear{Sneden}{1973}]{Sneden1973} Sneden 
C.~A., 1973, PhDT 


\bibitem[\protect\citeauthoryear{Stanek et al.}{1997}]{Stanek1997} 
Stanek K.~Z., Udalski A., Szymanski M., Kaluzny J., Kubiak M., Mateo M., 
Krzeminski W., 1997, ApJ, 477, 163 


\bibitem[\protect\citeauthoryear{Zoccali et 
al.}{2007}]{Zoccali2007} Zoccali M., et al., 2007, IAUS, 241, 73 


\bibitem[\protect\citeauthoryear{Zoccali et 
al.}{2008}]{Zoccali2008} Zoccali M., Hill V., Lecureur A., Barbuy B., Renzini A., Minniti D., G{\'o}mez A., Ortolani S., 2008, A\&A, 486, 177 


\end{thebibliography}
\end{document}